\documentclass[12pt]{article}

\usepackage{epsfig}

\usepackage{amssymb}
\usepackage{amsmath}
\usepackage{epsfig}
\usepackage{graphicx}

\def\be{\begin{equation}}
\def\ee{\end{equation}}
\def\ba{\begin{eqnarray}}
\def\ea{\end{eqnarray}}
\def\bea{\begin{eqnarray}}
\def\eea{\end{eqnarray}}
\def\mg{m_{g}}
\newcommand\de{\partial}
\newcommand{\sfrac}[2]{{\textstyle\frac{#1}{#2}}}

\begin{document}
\begin{titlepage}
\hskip 1 cm
\vskip 0.5cm

\vspace{25pt}
\begin{center}
    { \LARGE{\bf The New Ekpyrotic Ghost}}
    \vspace{33pt}

  {{\bf  Renata Kallosh,$^{1,2}$  Jin U Kang,$^{2,3}$ Andrei Linde$^{1,2}$ and Viatcheslav Mukhanov$^{2}$}}

    \vspace{20pt}

 {$^{1}$Department of Physics,
    Stanford University, Stanford, CA 94305, USA}

 {$^{2}$Arnold-Sommerfeld-Center for Theoretical Physics, Department f\"ur Physik,
Ludwig-Maximilians-Universit\"at M\"unchen,
Theresienstr. 37, D-80333, Munich, Germany}

{$^{3}$Department of Physics, Kim Il Sung University, Pyongyang, DPR. Korea}

  \end{center}

   \vspace{20pt}

\begin{abstract}
The new ekpyrotic scenario attempts to solve the singularity problem by involving violation of the null energy condition  in a model which combines the ekpyrotic/cyclic scenario with the ghost condensate theory  and the curvaton mechanism of production of adiabatic perturbations of metric. The Lagrangian of this theory, as well as of the ghost condensate model, contains a term with higher derivatives, which was added to the theory to stabilize its vacuum state. We found that this term  may affect the dynamics of the cosmological evolution.  Moreover, after a proper quantization, this term results in the existence of a new ghost field with negative energy, which leads to a catastrophic vacuum instability. We explain why one cannot treat this dangerous term as a correction valid only at small energies and momenta below some UV cut-off, and demonstrate the problems arising when one attempts to construct a UV completion of this theory.
\end{abstract}

\vspace{10pt}
\end{titlepage}

\tableofcontents

\section{Introduction: Inflation versus Ekpyrosis}

After more than 25 years of its development, inflationary theory gradually becomes a standard cosmological paradigm. It solves many difficult cosmological problems and makes several predictions, which  are in a very good agreement with observational data. There were many attempts to propose an alternative to inflation. In general, this could be a very healthy tendency. If one of these attempts will succeed, it will be of great importance. If none of them are successful, it will be an additional demonstration of the advantages of inflationary cosmology. However, since the stakes are high, we are witnessing a growing number of premature announcements of success in developing an alternative cosmological theory.

An instructive example is given by the ekpyrotic  scenario  \cite{KOST}. The authors of this scenario   claimed that it can solve all cosmological problems without using the stage of inflation. However, the original ekpyrotic scenario did not work. It is sufficient to say that the large mass and entropy of the universe remained unexplained, instead of solving the homogeneity problem this scenario only made it worse, and instead of the big bang expected in  \cite{KOST}, there was a big crunch  \cite{KKL,KKLTS}.

Soon after that, the ekpyrotic scenario was  replaced by the cyclic scenario, which used an infinite number of periods of expansion and contraction of the universe  \cite{cyclic}. Unfortunately, the origin of the scalar field potential required in this model, as well as in  \cite{KOST}, remains unclear, and the very existence of the cycles postulated in \cite{cyclic} has not been demonstrated.   When this scenario was analyzed  using the particular potential given in  \cite{cyclic} and  taking into account the effect of particle production in the early universe,  a very different cosmological regime was found \cite{Felder:2002jk,Linde:2002ws}.

The most difficult of the problems facing this scenario is the problem of the cosmological singularity. Originally there was a hope that the cosmological singularity problem will be solved in the context of string theory, but despite the attempts of the best experts in string theory, this problem remains unsolved  \cite{Liu:2002ft,Horowitz:2002mw,Berkooz}. Recently there were some developments in the analysis of this problem using the AdS/CFT correspondence \cite{Turok:2007ry}, but the results rely on certain conjectures and apply only to five dimensional space. As the authors admit, ``precise calculations are currently beyond reach'' for the physically interesting four dimensional space-time. This issue was previously studied  in \cite{Hertog:2005hu}, where it was concluded  that ``In our study of the field theory evolution, we find no evidence for a ÒbounceÓ from a big crunch to a big bang.''

In this paper we will discuss the recent development of this theory, called `the new ekpyrotic scenario' \cite{Creminelli:2006xe,Buchbinder:2007ad,Creminelli:2007aq,Buchbinder:2007tw}, which created a new wave of interest in the ekpyrotic/cyclic ideas. This is a rather complicated scenario, which attempts to solve the singularity problem by involving violation of the null energy condition (NEC) in a model which combines the ekpyrotic scenario \cite{KOST} with the ghost condensate theory  \cite{Arkani-Hamed:2003uy}  and the curvaton mechanism of production of adiabatic perturbations of metric \cite{curv,Lehners:2007ac}.

Usually the NEC violation leads to a vacuum instability, but the authors of \cite{Creminelli:2006xe,Buchbinder:2007ad,Creminelli:2007aq,Buchbinder:2007tw} argued that the instability occurs only near the bounce, so  it does not have enough time to fully develop. The instability is supposed to be dampened by higher derivative terms of the type $-(\square \phi)^2$ (the sign is important, see below), which were added to the action of the ghost condensate in \cite{Arkani-Hamed:2003uy}. These terms are absolutely essential in the new ekpyrotic theory for stabilization of the vacuum against the gradient and Jeans instabilities near the bounce.

However, these terms are quite problematic. Soon after introducing them, the authors of the ghost condensate theory, as well as several others, took a step back and argued that these terms cannot appear in any consistent theory, that the ghost condensate theory is ultraviolet-incomplete, that theories of this type lead to violation of the second law of thermodynamics,   allow construction of a {\it perpetuum mobile} of the 2nd kind, and therefore they  are incompatible with basic gravitational principles   \cite{Adams:2006sv,Dubovsky:2006vk,Eling:2007qd,ArkaniHamed:2007ky}.

These arguments did not discourage the authors of the new ekpyrotic theory and those who followed it, so we decided to analyze the situation in a more detailed way.  First of all, we found that the higher derivative terms were only partially taken into account in the investigation of perturbations, and were ignored in the investigation of the cosmological evolution in \cite{Creminelli:2006xe,Buchbinder:2007ad,Creminelli:2007aq,Buchbinder:2007tw}. Therefore the existence of  consistent and stable bouncing solutions postulated in the new ekpyrotic scenario required an additional investigation. We report the results of this investigation in Section \ref{bounce}.

More importantly, we found that these additional terms  lead to the existence of {\it new ghosts}, which have not been discussed in the ghost condensate theory and in the new ekpyrotic scenario \cite{Creminelli:2006xe,Buchbinder:2007ad,Creminelli:2007aq,Buchbinder:2007tw,Arkani-Hamed:2003uy}. In order to distinguish these ghosts from the relatively harmless condensed ghosts of the ghost condensate theory, we will call them {\it ekpyrotic ghosts}, even though, as we will show, they are already present in the ghost condensate theory. These ghosts lead to a catastrophic vacuum instability, quite independently of the cosmological evolution. In other words, the new ekpyrotic scenario, as well as the ghost condensate theory, appears to be physically inconsistent. But since  the new ekpyrotic scenario,  as different from the ghost condensate model, claims to solve the fundamental singularity problem by justifying the bounce solution, the existence of the ekpyrotic ghosts presents a much more serious problem for the new ekpyrotic scenario with such an ambitious goal. We describe this problem in Sections \ref{basic}, \ref{Ham}, \ref{Lagr}, \ref{EMT}, and \ref{instability}.

Finally, in Appendix we discuss certain attempts to save the new ekpyrotic scenario. One of such attempts is to say that this scenario is just an effective field theory which is valid only for sufficiently small values of frequencies and momenta. But then, of course, one cannot claim that this theory solves the singularity problem until its consistent UV completion with a stable vacuum is constructed. For example, we will show that if one simply ignores the higher derivative terms for frequencies and momenta above a certain cutoff, then the new ekpyrotic scenario fails to work because of the vacuum instability which is even much stronger than the ghost-related instability. We will also describe a possible procedure which may provide a consistent  UV completion of the theory with higher derivative terms of the type $+(\square \phi)^2$. Then we explain why this procedure fails for the ghost condensate and the new ekpyrotic theory where the sign of the higher derivative term must be negative.

\section{Ghost condensate and new ekpyrosis: The basic scenario}\label{basic}
The full description of the new ekpyrotic scenario is pretty involved. It includes two fields, one of which, $\phi$, is responsible for the ekpyrotic collapse, and another one, $\chi$, is responsible for generation of isocurvature perturbations, which eventually should be converted to adiabatic perturbations. Both fields must have quite complicated potentials, which can be found e.g. in \cite{Buchbinder:2007tw}. For the purposes of our discussion it is sufficient to consider a simplified model containing only one field, $\phi$. The simplest version of this scenario can be written as follows:
\begin{equation}
L=\sqrt{g}\left[M^{4}P(X)-{1\over 2} \left({\Box\phi\over M'}\right)^{2}\, -V(\phi)\right], \label{eq:General Lagrangian}\end{equation}
where $X=\frac{(\partial\phi)^{2}}{2m^{4}}$  is dimensionless. $P(X)$ is a dimensionless function which has a minimum at $X \not =0$. The first two terms in this theory represent the theory of a ghost condensate, the last one is the ekpyrotic potential. This potential is very small and very flat at large $\phi$, so for large $\phi$ this theory is reduced to the ghost condensate model of \cite{Arkani-Hamed:2003uy}.

The ghost condensate state corresponds to the minimum of $P(X)$. Without loss of generality one may assume that this minimum  occurs at $X = 1/2$, i.e. at $\partial_{i}\phi = 0$, $\dot \phi = -m^{2}$, so that $\phi = -m^{2}t$. As a simplest example, one can consider a function which looks as follows in the vicinity of its minimum:
\be
P(X) = {1\over 2}(X-1/2)^{2} \ .
\ee
The term $-{1\over 2} \bigl({\Box\phi\over M'}\bigr)^{2}$ was added to the Lagrangian in   \cite{Arkani-Hamed:2003uy} for stabilization of the fluctuations of the field $\phi$ in the vicinity of the background solution $\phi(t) = - m^{2} t$; more about it later.

This theory was represented in several different ways in \cite{Arkani-Hamed:2003uy,Creminelli:2006xe,Buchbinder:2007ad,Creminelli:2007aq,Buchbinder:2007tw}, where a set of parameters such as $K$ and  $\bar M   = M^{2}/M'$ was introduced.  The parameter $K$ can always be absorbed in a redefinition of $M$; in our notation, $K = 1$.

The equation for the homogeneous background can be represented
as follows:
\begin{equation}
\partial_{t}\left[a^{3}\left(P,_{X}\dot{\phi}+\frac{\partial_{t}(\ddot{\phi}+3H\dot{\phi})}{m_g^{2}}\right)\right]=-{a^{3}V,_{\phi}m^{4}\over M^{4}}\ , \label{eq:back ground EOM for simple case}\end{equation}
where we introduced the notation
\be
m_g   = \frac{M'M^{2}}{m^{2}} \ .
\ee
The meaning of this notation will be apparent soon.

The complete equation describing the dependence on the spatial coordinates
is \begin{equation}
\partial_{t}\left[a^{3}\left(P,_{X}\partial_{t}\phi+\frac{\partial_{t}(\Box\phi)}{m_g^{2}}\right)\right]-\partial_{i}\left[a\left(P,_{X}\partial_{i}\phi+\frac{\partial_{i}(\Box\phi)}{m_g^{2}}\right)\right]=-{a^{3}V,_{\phi}m^{4}\over M^{4}}.\label{eq:complete EOM}\end{equation}

Instead of solving these equations, the authors of   \cite{Creminelli:2006xe,Buchbinder:2007ad,Creminelli:2007aq,Buchbinder:2007tw}  analyzed (though not solved)  equation (\ref{eq:back ground EOM for simple case}) ignoring the higher derivative term $ {\partial_{t}(\ddot{\phi}+3H\dot{\phi})}/{m_g^{2}}$, assuming that it is small. Then they  analyzed  equation (\ref{eq:complete EOM}), applying it to  perturbations, ignoring the term $ {\partial_{t}(\Box\phi)}/{m_g^{2}}$, but keeping  the term $ {\partial_{i}(\Box\phi)}/{m_g^{2}}$, assuming that it is large. Our goal is to see what happens if one performs an investigation in a self-consistent way.

In order to do this, let us temporarily assume that the higher derivative term is absent, which corresponds to the limit $m_g\to \infty$. In this case our equation for $\phi$ reduces to the equation used in \cite{Creminelli:2006xe,Buchbinder:2007ad,Creminelli:2007aq,Buchbinder:2007tw}
 \begin{equation}
\partial_{t}[a^{3}P,_{X}\dot{\phi}]=-a^{3}V,_{\phi}m^{4}/M^{4} \ . \label{eq:EOM_0}\end{equation}
One of the Einstein equations, in the same approximation, is
\begin{equation}
 \dot{H}=-\frac{1}{2}(\varepsilon+p)=-M^{4}P,_{X}\, X= -M^{4}  X (X-1/2)\ ,\label{eq:Einstein EOM}
\end{equation}
where $\varepsilon$ is the energy density and $p$ is the pressure. (We are using the system of units where $M_p^2 =(8\pi G)^{-1} = 1$.)

The null energy condition (NEC) requires that $\varepsilon+p \geq 0$, and $\dot{H} \leq 0$. Therefore a collapsing universe  with $H<0$ cannot bounce back unless  NEC  is violated. It implies that the bounce can be possible only if $P,_{X}$  becomes negative, $P,_{X}< 0$, i.e. the field $X$ should become smaller than $1/2$.

It is convenient to represent the general solution for $\phi(t)$ as
\be
\phi(t) = -m^{2}t +\pi_{0}(t)+\pi(x_{i},t) \ ,
\ee
where $\pi_{0}(t)$ satisfies equation
\be
\ddot\pi_{0}+3H\dot \pi_{0} = -{m^{4}\over M^{4}} V,_{\phi} \ .
\ee
In this case one can show that the perturbations of the field $\pi(x_{i},t)$ have the following spectrum at small values of $P,_{X}$:
\be\label{badsp}
\omega^{2}  = P,_{X} k^{2} \ .
\ee
This means that $P,_{X}$ plays in this equation the same role as the square of the speed of sound.
 For small $P,_{X}$, one has
 \be
c_{s}^{2} = P,_{X} \ .
 \ee
The ghost condensate point $P,_{X} = 0$, which separates the region where NEC is satisfied and the region where it is violated, is the point where the perturbations are frozen. The real disaster happens when one crosses this border and goes to the region with $P,_{X} <0$, which corresponds to $c_{s}^{2} <0$. In this area the NEC is violated, and, simultaneously, perturbations start growing exponentially,
\be\label{instgr}
\pi_k(t) \sim  \exp({\sqrt {|c_{s}^{2}|}\, |k|\, t}) \sim \exp({\sqrt {|P,_{X}|}\, |k|\, t}) \ .
\ee
This is a disastrous gradient instability, which is much worse than the usual tachyonic instability. The tachyonic instability develops as $\exp (\sqrt{m^{2}-k^{2}}\, t)$, so its rate is limited by the tachyonic mass, and it occurs only for $k^{2} < m^{2}$. Meanwhile the   instability  (\ref{instgr}) occurs at all momenta $k$, and the rate of its development exponentially grows with the growth of $k$. This makes it abundantly clear how dangerous it is to  violate the null energy condition.

That is why it was necessary to add higher derivative terms of the type of $-{1\over 2} \left({\Box\phi\over M'}\right)^{2}$ to the ghost condensate Lagrangian \cite{Arkani-Hamed:2003uy}: The hope was that   such terms could provide at least some partial protection by changing the dispersion relation.

Since we are interested mostly in the high frequency effects corresponding to the rapidly developing instability, let us ignore for a while  the gravitational effects, which can be achieved by taking $a(t) = 1$, $H = 0$. In this case, the effective  Lagrangian for perturbations $\pi$ of the field $\phi$ in a vicinity of the minimum of $P(X)$ (i.e. for small $|P,_{X}|$) is
\be
L = {M^{4}\over m^{4}}\left[{1\over 2} \dot\pi^{2} - {1\over 2} P,_{X} (\nabla\pi)^{2} - {1\over 2 \mg^{2}} (\square \pi)^{2}\right] \ .
\label{Lp}
\ee
The equation of motion for the field $\pi$ is
\be\label{pip}
\ddot\pi  - P,_{X}\nabla^{2}\pi +  {1\over \mg^{2}}  \square^{2} \pi = 0\ .
\ee
At small frequencies $\omega$, which is the case analyzed in \cite{Arkani-Hamed:2003uy}, the dispersion relation corresponding to this  equation looks as follows:
\be\label{badspcorr}
\omega^{2}  = P,_{X} k^{2} +  {k^{4}\over m_g^{2}} \ .
\ee
This equation implies that the instability occurs only in some limited range of momenta $k$, which can be made small if the parameter $m_{g}$ is sufficiently small and, therefore, the higher derivative therm is sufficiently large. This is the one of the main assumptions of the new ekpyrotic scenario: If the violation of the NEC occurs only during a limited time near the bounce from the singularity, one can suppress the instability by adding a sufficiently large term $- {1\over 2 \mg^{2}} (\square \pi)^{2}$. (This term must have negative sign, because otherwise it does not protect us from the gradient instability. This will be important  for the discussion in Appendix.)

Note that one cannot simply add the higher derivative term and take it into account only up to some cut-off $\omega^{2}, k^{2} < \Lambda^{2}$. For example, if we ``turn on'' this term only at $k^{2} < \Lambda^{2}$, it is not going to save us from the gradient instability which occurs at $\omega^{2}  = P,_{X} k^{2}$ for all unlimitedly large $k$ in the region where the NEC is violated and $P,_{X} <0$.

There are several different problems associated with this scenario. First of all, in order to tame the instability during the bounce one should add a sufficiently large term $-{1\over 2} \bigl({\Box\phi\over M'}\bigr)^{2}$, which leads to the emergence of the term $ {1\over 2 \mg^{2}} (\square \pi)^{2}$ in the equation for $\pi$. But if this term is large, then one should not discard it in the equations for the homogeneous scalar field and in the Einstein equations, as it was done in \cite{Creminelli:2006xe,Buchbinder:2007ad,Creminelli:2007aq,Buchbinder:2007tw}.

The second problem is associated with the way the higher derivative terms were treated in \cite{Arkani-Hamed:2003uy,Creminelli:2006xe,Buchbinder:2007ad,Creminelli:2007aq,Buchbinder:2007tw}. The dispersion relation studied there was incomplete. The  full dispersion relation for the perturbations in the theory (\ref{Lp}), (\ref{pip}) is
\be\label{badspfull}
\omega^{2}  = P,_{X} k^{2} +   {(\omega^{2}-k^{2})^{2}\over m_g^{2}}\ .
\ee
This equation coincides with eq. (\ref{badspcorr}) in the limit of small $\omega$ studied in \cite{Arkani-Hamed:2003uy,Creminelli:2006xe,Buchbinder:2007ad,Creminelli:2007aq,Buchbinder:2007tw}. However,  this equation has two different branches of solutions, which we will present, for simplicity, for the case $P,_{X}=0$ corresponding to the minimum of the ghost condensate potential $P(X)$: \
\be
 \omega  = \pm\  \omega_{i} \ , \qquad i = 1,2 ,
\ee
where
\begin{equation} \label{spectrumfull1}
\omega_{1} = {1\over 2}\left(\sqrt { {m_g^{2} }+ 4 k^{2}}\, -\, {m_g }\right)  ,
\end{equation}
\begin{equation} \label{spectrumfull2}
\omega_{2} = {1\over 2}\left(\sqrt { {m_g^{2} }+ 4 k^{2}}\, +\, {m_g }\right)  .
\end{equation}
At high momenta, for $ k^{2} \gg {m_{g}^{2}}$, the spectrum for all 4 solutions is nearly the same
\begin{equation} \label{spectrumhigh}
\omega  \approx \pm |k| \ .
\end{equation}
At small momenta, for $ k^{2} \ll {m_{g}^{2}}$, one has two types of solutions: The lower frequency solution,  which was found in  \cite{Arkani-Hamed:2003uy}, is
\be
\omega  =\pm\  k^{2}/m_g \ .
\ee
But there is also  another, higher frequency solution,
\be \label{spectrumnew}
\omega  = \pm\  {m_g}  \ .
\ee

The reason for the existence of an additional branch of solutions is very simple. Equation for the field $\phi$ in the presence of the term with the higher derivatives is of the fourth order. To specify its solutions it is not sufficient to know the initial conditions for the field and its first derivative, one must know also the initial conditions for the second and the third derivatives. As a result, a single equation describes two different degrees of freedom.

To find a proper interpretation of these degrees of freedom, one must perform their quantization. This will be done in the next two sections. As we will show in these sections, the lower frequency solution corresponds to  normal particles with positive energy $\omega = +\omega_{1}(k)$, whereas the higher frequency solution corresponds to ekpyrotic ghosts with negative energy $-\omega_{2}(k)$.  The quantity $-{m_g}$ has the meaning of the ghost mass: it is given by the energy $\omega = -\omega_{2}(k)$ at  $k = 0$ and  it is negative.

\section{Hamiltonian quantization}\label{Ham}

 We see that our equations for $\omega$ have two sets of solutions, corresponding to states with positive and negative energy. As we will see now, some of them correspond to normal particles, some of them are ghosts. We will find below that the Hamiltonian based on the classical Lagrangian in eq. (\ref{Lp}) is
 \be
H_{quant} = \int {d^3k \over (2\pi)^3}
\left ( \omega_{1} a_k^\dagger a_k -\omega_{2} c_k^\dagger c_k\right ) \ .
\label{Hquant}\ee
The expressions for  $\omega_{1}$ and $\omega_{2}$ will be presented below for the case of generic $c_s^2$, for $c_s^2=0$ they are given in eqs. (\ref{spectrumfull1}) and (\ref{spectrumfull2}). Both  $\omega_{1}$ and $\omega_{2}$
are positive,  therefore $a_k^\dagger$ and $ a_k$ are creation/annihilation operators of normal particles whereas $c_k^\dagger$ and $ c_k$ are creation/annihilation operators of ghosts.

We will perform the quantization starting with the Lagrangian in  eq. (\ref{Lp}), with
 an arbitrary speed of sound,  $c_{s}^{2} = P,_{X}$. The case
 $c_{s}=1 $ is the Lorentz invariant Lagrangian. The case $c_{s}^{2} = P,_{X}=0 $ is the case considered in the previous section and appropriate to the ghost condensate and the new ekpyrotic scenario at the minimum of $P(X)$.

By rescaling the field $\pi \to {M^{2}\over m^{2}}\pi$ we have
\begin{equation}
L=\frac{1}{2}\left( \partial _{\mu }\pi \partial ^{\mu }\pi +\left(
c_{s}^{2}-1\right) \pi \Delta \pi -\frac{1}{\mg^{2}}(\square \pi )^{2}\right).
\label{1}
\end{equation}
This is the no-gravity theory considered in the  previous section. Note that the ghost condensate set up is already build in, the negative kinetic term for the original ghost  is eliminated by the condensate.
The existence of higher derivatives was only considered in \cite{Creminelli:2006xe,Buchbinder:2007ad,Creminelli:2007aq,Buchbinder:2007tw,Arkani-Hamed:2003uy} as a `cure' for the problem of stabilizing the system
after  the {\it original} ghost  condensation. As we argued in the previous section, this `cure' brings in  a {\it new ghost}, which remained unnoticed in \cite{Creminelli:2006xe,Buchbinder:2007ad,Creminelli:2007aq,Buchbinder:2007tw,Arkani-Hamed:2003uy}. In this section, as well as in the next one, we will present a detailed  derivation of this result.
Because of the presence of higher derivatives in the Lagrangian, the Hamiltonian quantization of this theory is somewhat nontrivial. It can be performed by the method invented by Ostrogradski \cite{Ostr}.

Thus we start with the rescaled eq. (\ref{Lp})
\be
L = {1\over 2}\left[ \dot\pi^{2} - c_s^2 (\nabla_x\pi)^{2} - {1\over  m_g^{2}} (\square \pi)^{2}\right] \ .
\label{Lresc}
\ee
The equation of motion for the field $\pi$ is
\be\label{eqm}
\ddot\pi  - c_s^2\nabla^{2}_x\pi +  {1\over \mg^{2}}  \square^{2} \pi = 0\ .
\ee
If the Lagrangian depends on the field $\pi$ and on its first  and second  time derivatives, the general procedure is the following. Starting with
$L= L( \pi, \dot \pi, \ddot \pi)$,
one defines 2 canonical degrees of freedom, $(q_1, p_1)$ and $(q_2, p_2)$:
\bea
&&q_1 \equiv \pi \ , \qquad p_1= {\partial L\over \partial \dot q_1} - {d\over dt} {\partial L\over \partial \ddot q_1} \ ,\nonumber\\
&&q_2 \equiv \dot \pi \ , \qquad p_2= {\partial L\over \partial \ddot q_1} \ .
\eea
The canonical Hamiltonian is
\be
H= p_1\dot q_1+p_2\dot q_2-L(q_1, q_2, \dot q_1, \dot q_2) \ .
\ee
The canonical Hamiltonian equations of motion,
\be
\dot q_{i} = {\partial H\over \partial p_{i}} \ ,   \qquad \dot p_{i} = -{\partial H\over \partial q_{i}} \ , \qquad i=1,2 \ ,
\label{Heq}\ee
are standard; they exactly reproduce the Lagrangian equation of motion (\ref{eqm}).
The quantization procedure requires  promoting the Poisson brackets to commutators which allows to identify the spectrum. There are many known examples of the Ostrogradski procedure of derivation of the canonical Hamiltonian, see for example \cite{de Urries:1998bi,Weldon:2003by}.

The Hamiltonian density constructed by the Ostrogradski procedure for the Lagrangian (\ref{Lresc})  is
\be
H_{cl} (\vec x, t) ={1\over 2}[ p_{1}^{2} -  (p_{1}-q_{2})^{2}-\mg^2 \left(p_{2} - {1\over \mg^2}  \nabla^{2}_{x}\, q_{1}  \right)^{2}  +c_s^2 (\nabla_x q_1)^{2}+ {1\over  \mg^{2} }\left(\nabla^{2}_{x}\, q_{1}  \right)^{2}] \ .
\ee
The next step in quantization is to consider the ansatz for the solution of classical equations of motion in the form
\be
q_1(\vec x, t) = \int {d^3 k \over (2\pi)^3} \left[{ f_k^1\over \sqrt{2\omega_{1}}} \; e^{-ik_1   x} + {f_k^2\over \sqrt{2\omega_{2}} } \; e^{ik_2   x} +\rm cc\right],
\label{cl} \ee
where $k_1 \,  x\equiv \omega_{1}( k)\,  t- \vec k \, \vec x$,  and $k_2 \, x\equiv \omega_{2}( k) \, t- \vec k \, \vec x$. We impose the Poisson brackets
\be
\{q_i, p_j\} = \delta_{ij}
\ee
and promote them to commutators of the type
\be
[q_i (\vec x, t), p_j (\vec x', t)] = i \delta_{ij} \delta^3(\vec x - \vec x') \ .
\label{com}\ee
This quantization condition requires  to promote the solution of the classical equation (\ref{cl}) to the quantum operator form, where
\be
f^1_k =  a_k  {  \mg  \over  \sqrt{\omega_{2}^2- \omega_{1}^2 }},  \qquad f^2_k =  c_k  { \mg \over   \sqrt{\omega_{2}^2- \omega_{1}^2 }} \ ,
\ee
and we impose normal commutation relation both on particles  with creation and annihilation operators $a^\dagger $ and $a$  and ghosts,  $c^\dagger$ and  $c$:
\be
[a_k, a^\dagger _{k'}] = (2\pi)^3 \delta^3 (\vec k -\vec k')\ , \qquad [c_k, c^\dagger _{k'}] = (2\pi)^3 \delta^3 (\vec k -\vec k') \ .
\ee
Here
 \be
 \omega_{1}(k^2; m_g, c_s^2) =\left( k^2+{m_g^2\over 2} - \sqrt{k^2m_g^2(1-c_s^2)+{m_g^4\over 4}}\right)^{1/2}
 \label{w1}\ee
and
\be
 \omega_{2}(k^2; m_g, c_s^2) =\left( k^2+{m_g^2\over 2} + \sqrt{k^2m_g^2(1-c_s^2)+{m_g^4\over 4}}\right)^{1/2}
  \label{w2}\ee
The Hamiltonian operator acquires a very simple form
\be
H_{quant} = {1\over 2} \int {d^3k \over (2\pi)^3}
\left ( \omega_{1} ( a_k^\dagger a_k +a_k a_k^\dagger ) -\omega_{2} ( c_k^\dagger c_k+ c_kc_k^\dagger ) \right )= \int {d^3k \over (2\pi)^3}
\left ( \omega_{1} a_k^\dagger a_k -\omega_{2} c_k^\dagger c_k+C\right )  \ .
\ee
Here the infinite term  $C$ is equal to $ {(2\pi)^3\over 2} (\omega_{1}-\omega_{2})\delta^3(0)$.  It represents the infinite shift of the vacuum energy due to the sum of all modes of the zero-point energies and is usually neglected in  quantum field theory.
Apart  from this  infinite $c$-number this is an expression promised in eq. (\ref{Hquant}).

We now define the vacuum state $ |0\rangle$ as the state which is annihilated both by  the particle as well as by the  ghosts annihilation operators, $ a_k  |0\rangle= c_k  |0\rangle= 0$.  Thus the energy operator acting on a state of a particle   has  positive  value and on a state of a ghost  has the negative  value\footnote{One could use an alternative way of ghost quantization, by changing the sign of their commutator relations. In this case the ghosts would have positive energy, but this would occur at the expense of introducing a nonsensical notion of negative probabilities.}
\be
H_{quant} \; a_k^\dagger |0\rangle =  \omega_{1}(k) \; a_k^\dagger |0\rangle\ , \qquad
H_{quant} \; c_k^\dagger |0\rangle = - \omega_{2}(k) \; c_k^\dagger |0\rangle \ .
\ee
This confirms the physical picture outlined at the end of the previous section.

\section{Lagrangian quantization}\label{Lagr}
 The advantage of the Hamiltonian method is that it gives an unambiguous definition of the quantum-mechanical energy operator, which is negative for ghosts. This is most important for our subsequent analysis of the vacuum instability in the new ekpyrotic scenario. However, it is also quite  instructive to explain the existence of the ghost field in new ekpyrotic  scenario using  the Lagrangian approach. The Lagrangian formulation is very convenient for coupling of the model to gravity.

 Using
Lagrangian multiplier,  one can  rewrite eq. (\ref{1}) as
\begin{equation}
L=\frac{1}{2}\left( \partial _{\mu }\pi \partial ^{\mu }\pi +\left(
c_{s}^{2}-1\right) \pi \Delta \pi -\frac{B^{2}}{\mg^{2}}\right) +\lambda
\left( B-\square \pi \right) .  \label{2}
\end{equation}%
Variation with respect to $B$ gives $
\lambda =\frac{B}{m_{g}^{2}}$.
After substituting $\lambda $ in (\ref{2}) and skipping total derivative we
obtain%
\begin{eqnarray}
L &=&\frac{1}{2}\left( \partial _{\mu }\pi \partial ^{\mu }\pi +\left(
c_{s}^{2}-1\right) \pi \Delta \pi +\frac{B^{2}}{\mg^{2}}\right) +\frac{1}{\mg^{2}%
}\partial _{\mu }B\,\partial ^{\mu }\pi   \notag \\
&=&\frac{1}{2}\partial _{\mu }\left( \pi +\frac{2B}{\mg^{2}}\right)
\partial ^{\mu }\pi +\frac{B^{2}}{2\mg^{2}}+\frac{1}{2}\left(
c_{s}^{2}-1\right) \pi \Delta \pi \ .  \label{3}
\end{eqnarray}%
Introducing new variables $\sigma,\ \xi$ according to
\begin{eqnarray*}
\sigma+\xi &=&\pi +\frac{2B}{\mg^{2}}, \\
\sigma-\xi &=&\pi \ ,
\end{eqnarray*}%
and substituting $\pi =\sigma-\xi,$ $B=\square \pi= \mg^{2}\xi$ in (\ref{3}) we obtain
\begin{equation}
L=\frac{1}{2}\left( \partial _{\mu }\sigma\partial ^{\mu }\sigma-\partial
_{\mu }\xi\partial ^{\mu }\xi+\mg^{2}\xi^{2}\right) +\frac{1}{2}\left(
c_{s}^{2}-1\right) (\sigma-\xi)\Delta (\sigma-\xi) \ . \label{4}
\end{equation}
In the case $c_{s}^{2}=1$ we have two decoupled scalar fields: massive with
negative kinetic energy and massless with positive kinetic energy.
\begin{equation}
L_{c_{s}^{2}=1}=\frac{1}{2}\left( \partial _{\mu }\sigma\partial ^{\mu }\sigma-\partial
_{\mu }\xi\partial ^{\mu }\xi+\mg^{2}\xi^{2}\right)  \ . \label{nocs}
\end{equation}%

For the homogeneous field ($k=0$ mode) the Lagrangian does not depend on $c_s^2$ and is reduced to
\begin{equation}
L_{k=0}=\frac{1}{2}\left( \dot\sigma^2-\dot \xi^2+\mg^{2}\xi^{2}\right)  \ . \label{nocs2}
\end{equation}%
The relation between the $k=0$ mode of the original field $\pi$, the normal field $\sigma$ and the ghost field $\xi$ is
\be\label{defghost}
 \sigma = \pi+ \xi \ ,\qquad \xi   =  {\ddot \pi\over \mg^2} \ .
\ee

When
$c_{s}^{2}\neq 1$ and $k\neq 0$ these fields still couple.  To
diagonalize the Lagrangian in eq.  (\ref{1}) and decouple the oscillators we have to go to  normal
coordinates, similar to the case of the classical mechanics  of coupled
harmonic oscillators. For that we need to solve the eigenvalue problem and
find the eigenfrequencies of the oscillators. Let us consider the modes with the
wavenumbers $\mathbf{k.}$ For such modes we can perform the following change of variables
\be
 \sigma_{\mathbf{k}} \equiv  {\ddot \pi_{\mathbf{k}} +\omega_{2}^2\pi_{\mathbf{k}} \over m_g\sqrt{\omega_{2}^2-\omega_{1}^2}}\ , \qquad  \xi_{\mathbf{k}} \equiv {\ddot \pi_{\mathbf{k}} +\omega_{1}^2\pi_{\mathbf{k}} \over m_g\sqrt{\omega_{2}^2-\omega_{1}^2}} \ ,
 \ee
 where $\omega_{1}, \omega_{2}$ are defined in eqs. (\ref{w1}), (\ref{w2}). In the special case of $c_{s}^{2}=0$ the answers for $\omega_{1}, \omega_{2}$ simplify and are shown in eqs. (\ref{spectrumfull1}) and (\ref{spectrumfull2}).
After a change of variables   we find for these modes  in the momentum space
 \begin{equation}
\tilde L_{c_{s}^{2}} =\frac{1}{2}\left( \overset{\boldsymbol{.}}{{\sigma}}_{\mathbf{k}}\overset{%
\boldsymbol{.}}{{\sigma}}_{-\mathbf{k}}-\omega _{1}^{2}\,{\sigma}_{\mathbf{k}%
}{\sigma}_{-\mathbf{k}}-\overset{\boldsymbol{.}}{{ \xi}}_{\mathbf{k}}%
\overset{\boldsymbol{.}}{{ \xi}}_{-\mathbf{k}}+\omega _{2}^{2}\,{ \xi}_{%
\mathbf{k}}{\xi}_{-\mathbf{k}}\right).
\label{Slava}\end{equation}
The modes of $\sigma$ are normal, and the modes of $\xi$ are ghosts.
Using this Lagrangian, one can easily confirm the  final result of the Hamiltonian quantization given in the previous section.\footnote{After we finished this paper, we learned that the Lagrangian quantization of the ghost condensate scenario was earlier performed by Aref'eva and Volovich for the case $c_{s}^{2} = P,_{X}=0 $ \cite{Aref'eva:2006xy}, and they also concluded that this scenario suffers from the existence of ghosts. When our works overlap, our results agree with each other. We use the Lagrangian approach mainly to have an alternative derivation of the results of the Hamiltonian quantization. The Hamiltonian approach clearly establishes the energy operator and the sign of its eigenvalues, which is necessary to have an unambiguous proof  that the energy of the ghosts is indeed negative and the ghosts do not disappear at non-vanishing $c_{s}^{2} = P,_{X}\neq 0 $ when the ekpyrotic universe is out of the ghost condensate minimum.} The classical mode $\sigma_{\mathbf{k}} $ is associated with creation/annihilation operators $a^{\dagger}_k, a_k $ of normal particles after quantization   and the classical mode  $ \xi_{\mathbf{k}}$ is associated with creation/annihilation operators of ghosts particles $c^{\dagger}_k, c_k $ after quantization. A quantization of the theory in eq. (\ref{Slava}) leads to the Hamiltonian in eq. (\ref{Hquant}).

\section{Energy-momentum tensor and equations of motion}\label{EMT}

First, we will compute the energy-momentum tensor (EMT) of the Lagrangian (\ref{4}) using the Noether procedure:\[
T_{\nu}^{\mu}=\frac{\partial L}{\partial(\partial_{\mu}\varphi)}\partial_{\nu}\varphi-L\,\eta_{\nu}^{\mu},\]
 where $\eta_{\nu}^{\mu}$ is Minkowski metric.
We find  \begin{eqnarray}
T_{\mu\nu} & = & \partial_{\mu}\sigma\partial_{\nu}\sigma-\partial_{\mu}\xi\partial_{\nu}\xi+\eta_{\mu i}(c_{s}^{2}-1)\partial^{i}(\sigma-\xi)\partial_{\nu}(\sigma-\xi)\nonumber \\
 &  & -\eta_{\mu\nu}\left(\frac{1}{2}(\partial_{\alpha}\sigma\partial^{\alpha}\sigma-\partial_{\alpha}\xi\partial^{\alpha}\xi+
 m_{g}^{2}\xi^{2})+\frac{1}{2}(c_{s}^{2}-1)\partial_{i}(\sigma-\xi)\partial^{i}(\sigma-\xi)\right).\label{eq:EMT for arbitrary c_s^2}\end{eqnarray}
 The energy density is \[
\varepsilon=T_{00}=\frac{1}{2} [ \dot{\sigma}^{2}+ (\partial_{i}\sigma)^2-\dot{\xi}^{2}-
(\partial_{i}\xi)^2 -m_{g}^{2}\xi^{2}+(c_{s}^{2}-1)(\partial_{i}(\sigma-\xi))^2] .\]

For the homogeneous field ($k=0$ mode), the energy density
can be split into two parts, i.e. a normal field part and an ekpyrotic ghost field part:\[
\varepsilon=\varepsilon_{\sigma}+\varepsilon_{\xi},\]
 where \[
\varepsilon_{\sigma}=\frac{1}{2}\dot{\sigma}^{2}>0,\,\qquad \varepsilon_{\xi}=-\frac{1}{2}\dot{\xi}^{2}-\frac{1}{2}m_{g}^{2}\xi^{2}<0\ .\]
 Thus the energy of the ghost field $\xi$ is negative.

Up to now we have turned off gravity. In the presence of gravity,
the energy-momentum tensor of the full Lagrangian (1) in Sec. 2 is
calculated by varying the action with respect to the metric:\begin{eqnarray}
T_{\mu\nu} & = & g_{\mu\nu}\left[-M^{4}P(X)-\frac{(\Box\phi)^{2}}{2M'^{2}}+V(\phi)-\frac{\partial_{\alpha}(\Box\phi)\partial^{\alpha}\phi}{M'^{2}}\right]\nonumber \\
 &  & +M^{4}m^{-4}P,_{X}\partial_{\mu}\phi\partial_{\nu}\phi+M'^{-2}(\partial_{\mu}(\Box\phi)\,\partial_{\nu}\phi+\partial_{\nu}(\Box\phi)\partial_{\mu}\phi)\nonumber \\
 & \equiv & g_{\mu\nu}\left[-M^{4}P(X)-\frac{M'^{2}Y^{2}}{2}+V(\phi)-\partial_{\alpha}Y\,\partial^{\alpha}\phi\right]\nonumber \\
 &  & +M^{4}m^{-4}P,_{X}\partial_{\mu}\phi\partial_{\nu}\phi+\partial_{\mu}Y\,\partial_{\nu}\phi+\partial_{\nu}Y\,\partial_{\mu}\phi \ ,\label{eq:EMT for phi}\end{eqnarray}
 where
 \be
 Y\equiv M'^{-2}\Box\phi \ .
 \ee

From this, for a homogeneous, spatially flat FRW space time we have
the energy density \begin{equation}
\varepsilon=M^{4}(2P,_{X}\, X-P(X))+V(\phi)-\frac{M'^{2}Y^{2}}{2}+\dot{Y}\,\dot{\phi}\label{eq:energy density}\end{equation}
 and the pressure \begin{equation}
p=M^{4}P(X)-V(\phi)+\frac{M'^{2}Y^{2}}{2}+\dot{Y}\,\dot{\phi}\ ,\label{eq:pressure}\end{equation}
 so that \begin{equation}
\dot{H}=-\frac{1}{2}(\varepsilon+p)=-M^{4}P,_{X}\, X-\dot{Y}\dot{\,\phi} \ .\label{eq:Einstein EOM2}\end{equation}

Note that in the homogeneous case in the absence of gravity the ekpyrotic ghost field $\xi$ as defined in eq. (\ref{defghost}) is directly proportional to the field $Y$:
\be
\xi = {m^{2}\over M^{2}}\  Y \ .
\ee

The closed equations of motion, which we used for our numerical analysis,
are obtained as follows:\begin{eqnarray*}
\ddot{\phi}\left(P,_{X}+2X P,_{XX}\right)+3H\, P,_{X}\,\dot{\phi}+\frac{m^{4}}{M^{4}}(\ddot{Y}+3H\dot{Y}) & = & -V,_{\phi}m^{4}/M^{4} \ ,\end{eqnarray*}
 \begin{equation}
\dot{H}=-M^{4}P,_{X}\,\frac{\dot{\phi}^{2}}{2m^{4}}-\dot{Y}\dot{\,\phi} \ , \label{eq:closed EOM for phi}\end{equation}
 \[
M'^{-2}(\ddot{\phi}+3H\dot{\phi})=Y \ . \]
 Here $X=\dot{\phi}^{2}/2m^{4}$.

In these equations the higher derivative corrections appear in the
terms containing the derivatives of $Y$. The last of these equations shows that $Y\to0$
in the limit $M'\to\infty$ (i.e. $m_{g}\to\infty$), and then the
dynamics reduces to one with no higher derivative corrections.

The closed equations of motion for $\pi$ coupled to gravity are obtained
by expanding \eqref{eq:closed EOM for phi} and linearizing with respect
to $\pi$ and $Y$. Then we get \[
\ddot{\pi}+3H\,\dot{\pi}+\frac{m^{4}}{M^{4}}\,(\ddot{Y}+3H\dot{\, Y})=-V,_{\phi}m^{4}/M^{4}\ , \]
 \begin{equation}
\ddot{\pi}+3H\,\dot{\pi}=M'^{2}\, Y+3H\, m^{2}\ , \label{eq:linearized EOM for pi}\end{equation}
 \[
\dot{H}=\frac{M^{4}\dot{\pi}}{2m^{2}}+\dot{Y}\, m^{2}\ .\]

\section{On reality of the bounce and reality of ghosts}\label{bounce}
Using the equations derived above, we performed an analytical and  numerical investigation of the possibility of the bounce in the new ekpyrotic scenario. We will not present all of the details of this investigation here since it contains a lot of material which may distract the reader  from the main conclusion of our paper,  discussed in the next section: Because of the existence of the ghosts, this theory suffers from a catastrophic vacuum instability. If this is correct,  any analysis of  classical dynamics has very limited significance. However, we will briefly discuss  our main findings here, just to compare them with the expectations expressed in \cite{Creminelli:2006xe,Buchbinder:2007ad,Creminelli:2007aq,Buchbinder:2007tw}.

Our investigation was based on the particular scenario discussed in \cite{Buchbinder:2007ad,Buchbinder:2007tw} because no explicit form of the full ekpyrotic potential was presented in \cite{Creminelli:2006xe,Creminelli:2007aq}. The authors of  \cite{Buchbinder:2007ad,Buchbinder:2007tw} presented the full ekpyrotic potential, but they did not fully verify the validity of their scenario, even in the absence of the higher derivative terms.

Before discussing our results taking into account higher derivatives, let us remember several constraints on the model parameters which were derived in \cite{Creminelli:2006xe,Buchbinder:2007ad,Creminelli:2007aq,Buchbinder:2007tw}. We will represent these constraints in terms of the ghost condensate mass instead of the parameter $M'$, for $K = 1$. In this case the  stability condition (7.19) in \cite{Buchbinder:2007ad} (see also  \cite{Creminelli:2006xe,Creminelli:2007aq}) reads:
\be\label{instab}
{|\dot H|\over |H|} \lesssim {  M^{4} \over m_{g}} \lesssim |H| \ .
\ee
It was assumed in \cite{Buchbinder:2007ad} that the bounce should occur very quickly, during the time $\Delta t \lesssim |H_{0}|^{{-1}} \sim 1/ \sqrt{p |V_{min}|}$. Here $H_{0}$ is the Hubble constant at the end of the ekpyrotic state, just before it start decreasing during the bounce, $p \sim 10^{{-2}}$, and $V_{min}$ is the value of the ekpyrotic potential in its minimum. During the bounce one can estimate   $\Delta H \sim |H_{0}| \sim \dot H \Delta t \lesssim {\dot H\over |H_{0}|}$ because we assume, following  \cite{Buchbinder:2007ad}, that $\Delta t < |H_{0}|^{{-1}}$, and we assume an approximately linear change of $H$ from $-|H_{0}|$ to $|H_{0}|$. This means that $ {\dot H\over |H_{0}|} \gtrsim |H_{0}|$. In this case the previous inequalities become quite restrictive,
\be\label{instab2}
|H_{0}| \lesssim  {  M^{4} \over m_{g}} \lesssim |H_{0}| \ .
\ee
This set of inequalities requires that the stable bounce is not generic; it can occur only for a fine-tuned value of the ghost mass,
\be\label{instab3}
m_{g}  \sim  {  M^{4} \over |H_{0}|} \sim {M^{4}\over\sqrt{p |V_{min}|}} \ .
\ee
The method of derivation of these conditions required an additional condition to be satisfied,
 $|H_{0}| \sim \sqrt{p |V_{min}|} \ll  {  M^{2} }$, see Eqs. (8.8) and (8.17) of Ref. \cite{Buchbinder:2007ad}. This condition is satisfied for
\be
m_{g} \gg M^{2} \ .
\ee

Whereas the condition (\ref{instab3}) seems necessary in order to avoid the development of the gravitational instability and the gradient instability during the bounce for $m_{g}\gg M^{2}$, it is not sufficient, simply because the very existence of the bounce may require $m_{g}$ to be very much different from its fine-tuned value $m_{g}  \sim  {M^{4}\over\sqrt{p |V_{min}|}}$. 
 \begin{figure}[ht!]
\centering{\hskip 0.5cm
\includegraphics[height=4.5cm]{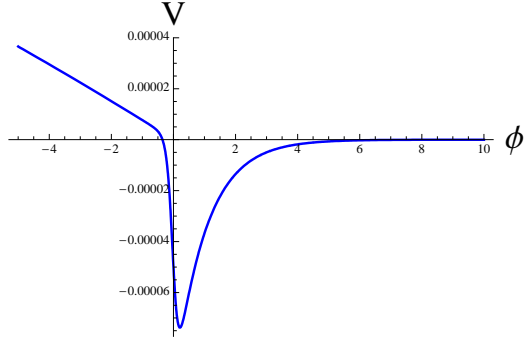}}
\caption{\small The ``new ekpyrotic potential,'' see Fig. 3 in \cite{Buchbinder:2007ad} and Fig. 6 in \cite{Buchbinder:2007tw}. The cosmological evolution in this model results in a universe with a permanently growing rate of expansion after the bounce, which is unacceptable.} \label{fig:Figp1}
\end{figure}
Indeed, our investigation of the cosmological evolution in this model shows that generically  the bounce does not appear at all, or one encounters a singular behavior of $\ddot\phi$  because of the vanishing of the term $P,_{X}+2X P,_{XX}$ in (\ref{eq:closed EOM for phi}), or one finds an unstable bounce, or the bounce ends up with an unlimited growth of the Hubble constant, like in the Big Rip scenario \cite{McInnes:2001zw}. Finding a proper potential leading to a desirable cosmological evolution requires a lot of fine-tuning, in addition to the fine-tuning already described in \cite{Buchbinder:2007ad,Buchbinder:2007tw}.

\begin{figure}[ht!]
\centering{\hskip 0.5cm
\includegraphics[height=5cm]{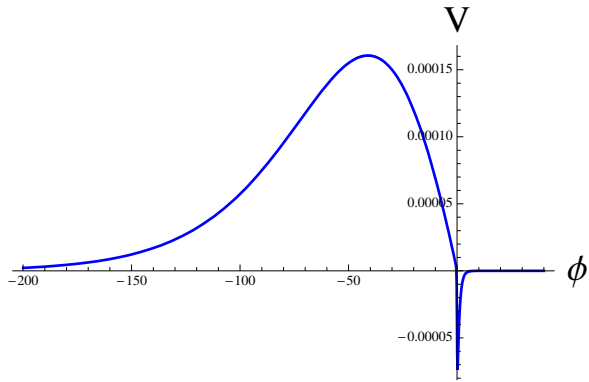}}
\caption{\small An improved potential which  leads to a bounce followed by a normal cosmological evolution. We do not know whether this extremely fine-tuned potential can be derived from any realistic theory.} \label{fig:Figp2}
\end{figure}

\begin{figure}[ht!]
\centering{\hskip 0.5cm
\includegraphics[height=4.5cm]{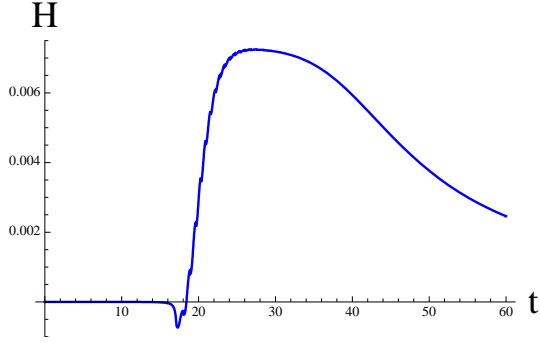}}
\caption{\small The behavior of the Hubble constant $H(t)$ near the bounce, which occurs near $t = 18$. To verify the stability of the universe during the bounce, one would need to perform an additional investigation taking into account the ghost field oscillations shown in Fig. \ref{fig:Figb2}.} \label{fig:Figb1}
\end{figure}

For example, the bounce in the model with the ``new ekpyrotic potential'' described in \cite{Buchbinder:2007ad,Buchbinder:2007tw} and shown in Fig. \ref{fig:Figp1} results in a universe with a permanently growing rate of expansion after the bounce, which would be absolutely different from our universe. To avoid this disaster, one must bend the potential, to make it approaching the value corresponding to the present value of the cosmological constant, see Fig. \ref{fig:Figp2}. This bending should not be too sharp, and it should not begin too early, since otherwise the universe bounces back and ends up in the singularity. Fig. \ref{fig:Figb1} shows the bouncing solution in the theory with this potential.
\begin{figure}[ht!]
\centering{\hskip 0.5cm
\includegraphics[height=6cm]{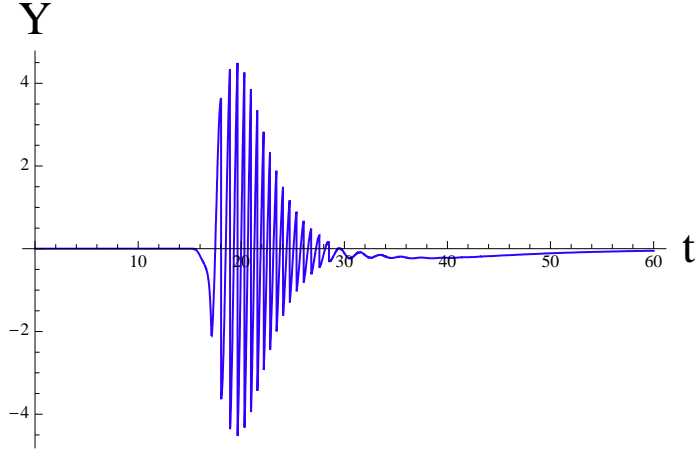}}
\caption{\small Ekpyrotic ghost field oscillations.} \label{fig:Figb2}
\end{figure}

Our calculations clearly demonstrate the reality of the ekpyrotic ghosts, see Fig.  \ref{fig:Figb2}, which shows the behavior of the ghost-related field $Y = {M^{2}\over m^{2}}\xi$ near the bounce.
The oscillations shown in Fig.  \ref{fig:Figb2} represent the ghost matter with negative energy, which was generated during the ekpyrotic collapse. We started with initial conditions  $Y=\dot Y = 0$, i.e. in the vacuum without ghosts, and yet the ghost-related  field $Y$ emerged dynamically. It oscillates with the frequency which is much higher than the rate of the change of the average value of the field $\phi$.

This shows that the ekpyrotic ghost is not just a mathematical construct or a figment of imagination, but a real field.
We have found  that the amplitude of the oscillations of the ghost field is very sensitive to the choice of initial conditions; it may be negligibly small or very large. Therefore in the investigation of the cosmological dynamics one should not simply consider the universe filled with scalar fields or scalar particles. The universe generically will contain normal particles and ghost particles and fields with negative energy. The ghost particles will interact with normal particles in a very unusual way:  particles and ghosts will run after each other with ever growing speed. This regime is possible because when the normal particles gain energy, the ghosts loose energy, so  the  acceleration regime is consistent with energy conservation. This unusual instability, which is very similar to the process to be considered in the next section,  can make it especially difficult to solve the homogeneity problem in this scenario.

\section{Ghosts, singularity and vacuum instability}\label{instability}

It was not the goal of the previous section to prove that  the ghosts do not allow one to  solve the singularity problem. They may or may not  spoil the bounce in the new ekpyrotic scenario. However,  in general, if one is allowed to introduce ghosts, then the solution of the singularity problem becomes nearly trivial, and it does not require the ekpyrotic scenario or the ghost condensate.

Indeed, let us consider a simple model describing a flat collapsing universe which contains a dust of heavy non-relativistic particles with initial energy density $\rho_{M} $, and a gas of ultra-relativistic ghosts with  initial energy density $-\rho_{g}  < 0$. Suppose that   at the initial moment  $t=0$, when the scale factor of the universe was equal to $a(0) = 1$, the energy density was dominated by energy density of normal particles, $\rho_{M}-\rho_{g} > 0$. The absolute value of the ghost energy density in the collapsing universe grows faster than the energy of the non-relativistic matter.  The Friedmann equation describing a collapsing universe is
\be
H^{2 }= \left ({\dot a\over a}\right)^{2} =  {\rho_{M}\over a^{3}} -  {\rho_{g}\over a^{4}} \ .
\ee
 In the beginning of the cosmological evolution, the universe is collapsing, but when the scale factor shrinks to $a_{\rm bounce} = {\rho_{g}\over \rho_{M}}$, the  Hubble constant vanishes, and  the universe bounces back, thus avoiding the singularity.

 Thus nothing can be easier than solving the singularity problem once we invoke  ghosts to help us in this endeavor, unless we are worried about the gravitational instability problem mentioned in the previous section. Other examples of the situations when ghosts save us from the singularity can be found, e.g. in \cite{Cai:2007zv}, where the authors not only  study a way to avoid the singularity with the help of ghosts, but even investigate the evolution of metric perturbations  during the bounce.  So what can be wrong with it?

A long time ago, an obvious answer would be that  theories with ghosts lead to negative probabilities,  violate unitarity and therefore do not make any sense whatsoever. Later on, it was realized that if one treats ghosts as particles with negative energy, then problems with unitarity are replaced by the problem of vacuum stability due to interactions between ghosts and normal particles with positive energy, see, e.g. \cite{Linde:1984ir,Linde:1988ws,Carroll:2003st,Cline:2003gs,Buniy:2005vh,Kaplan:2005rr,Buniy:2006xf}. Indeed, unless the ghosts are hidden in another universe \cite{Linde:1988ws}, nothing can forbid creation of pairs of ghosts and normal particles under the condition that their total momentum and energy vanish. Since the total energy of ghosts is negative, this condition is easy to satisfy.

\begin{figure}[ht!]
\centering{\hskip 0.5cm
\includegraphics[height=5.5cm]{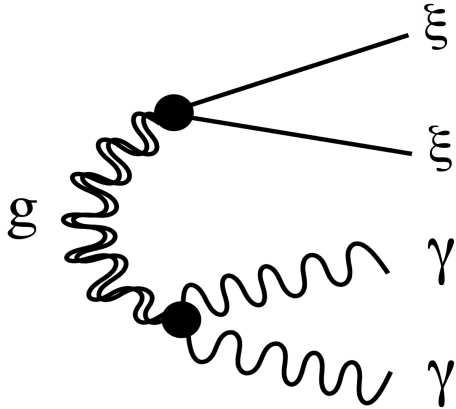}}
\caption{Vacuum decay with production of ghosts $\xi$ and usual particles $\gamma$ interacting with each other by the graviton exchange.} \label{fig:creation}
\end{figure}

There are many channels of vacuum decay; the simplest and absolutely unavoidable one is due to the universal gravitational interaction between ghosts and all other particles, e.g. photons. An example of this interaction was considered in \cite{Cline:2003gs}, see Fig. \ref{fig:creation}. Nothing can forbid this process because it does not require any energy input: the positive energy of normal particles can be  compensated by the negative energy of ghosts.

 An investigation of the rate of the vacuum decay in this process leads to a double-divergent result. First of all, there
is a power-law divergence  because nothing forbids creation of particles with indefinitely large energy. In addition, there is also a quadratic divergence in the integral over velocity  \cite{Cline:2003gs,Kaplan:2005rr}. This leads to  a catastrophic vacuum decay.

Of course, one can always argue that such processes are impossible or suppressed because of some kind of cutoff in momentum space, or further corrections, or non-local interactions. However, the necessity of introducing such a cut-off, or additional corrections to corrections, after introducing the higher derivative terms which were supposed to work as a cutoff in the first place, adds a lot to the already very high price of proposing  an alternative to inflation: First it was the ekpyrotic theory, then the ghost condensate and curvatons, and finally - ekpyrotic ghosts with negative energy which lead to a catastrophic vacuum instability. And if we are ready to introduce an ultraviolet  cutoff in momentum space, which corresponds to a small-scale cutoff in space-time, then why would we even worry about the singularity problem, which is supposed to occur on an infinitesimally small space-time scale?

In fact, this problem was already emphasized by the authors of the new ekpyrotic scenario, who wrote \cite{Buchbinder:2007ad}:

``But ghosts have disastrous consequences for the viability of the theory. In order to regulate the rate of vacuum decay one must invoke explicit Lorentz breaking at some low scale~\cite{Cline:2003gs}. In any case there is no sense in which a theory with ghosts can be thought as an effective theory, since the ghost instability is present all the way to the UV cut-off of the theory.''

We have nothing to add to this characterization of their own model.

\

\noindent {\bf Acknowledgments}:  It is a pleasure to thank  B.~Craps, P.~Creminelli, G.~Horowitz, N.~Kaloper, J.~Khoury,  S.~Mukohyama,   L.~Senatore, V.~Vanchurin,  A.~Vikman and S.~Winitzki for useful discussions. The work by R.K. and A.L. was supported in part by NSF grant PHY-0244728 and by the Alexander-von-Humboldt Foundation. The work by J.U.K. was supported by the German Academic Exchange Service.

\

\section{Appendix. Exorcising ghosts?}\label{higher}

After this paper was submitted, one of the authors of the new ekpyrotic scenario argued \cite{Khoury} that, according to \cite{Creminelli:2005qk}, ghosts can be removed by field redefinitions and adding other degrees of freedom in the effective UV theory \cite{Creminelli:2005qk}. Let us reproduce this argument  and  explain why it does not apply to the ghost condensate theory and to the new ekpyrotic scenario.

Refs. \cite{Khoury,Creminelli:2005qk} considered a normal
massless scalar field $\phi$ with Lagrangian density in  $(-,+,+,+)$ signature.\footnote{In our paper we used the signature $(+,-,-,-)$, so some care should be taken when comparing the equations. Note that this does not change the sign of the higher derivative term $\sim a(\Box \phi)^2$; the ghost condensate/ekpyrotic  theory corresponds to $a = -1$.}
\be \label{4d_L}
{\cal L} = -\sfrac12(\de \phi)^2 + \frac a {2 m_g^2} (\Box \phi)^2
- V_{\rm int}(\phi)\; ,
\ee
where 
$a=\pm 1$, and $V_{\rm int}$ is a self-interaction term.  This theory is similar to the ghost condensate/new ekpyrotic  theory in the case $a = -1$, $c_{s} = 1$,  see eqs. (1) and (\ref{1}). The sign of $a$ is crucially important: the term $+\frac 1 {2 m_g^2} (\Box \phi)^2$ would not protect this theory against the gradient instability in the region with the NEC violation.

Note that in notation of   \cite{Khoury,Creminelli:2005qk}, $m_{g}= \Lambda$, which could suggest that the ghost mass is a UV cut-off, and therefore there are no dangerous excitations with energies and momenta higher than $m_{g}$.  However, this interpretation of the theory (\ref{4d_L})  would be  misleading.  Upon a correct quantization, this theory  can be represented as a theory of two fields  without the higher derivative non-renormalizable term $ \frac a {2 m_g^2} (\Box \phi)^2$, see Eq. (\ref{4}).   One can introduce the UV cut-off $\Lambda$ when regularizing  Feynman diagrams in this theory, but there is absolutely no reason to identify it with $m_{g}$; in fact, the UV cut-off which appears in the regularization procedure is supposed to be arbitrarily large, so the perturbations with frequencies greater than $m_{g}$ should not be forbidden.

Moreover, as we already explained in Section 2,  one cannot take the higher derivative term into account only up to some cut-off $\omega^{2}, k^{2} < \Lambda^{2}$. If, for example, we ``turn on'' this term only at $k^{2} < \Lambda^{2}$, it is not going to protect us from the gradient instability, which occurs at $\omega^{2}  = P,_{X} k^{2}$ for all indefinitely large $k$ in the region where the NEC is violated and $P,_{X} <0$. Note that this instability grows stronger for greater values of momenta $k$. Therefore if one wants to prove that the new ekpyrotic scenario does not lead to instabilities, one must verify it for {\it all} values of momenta. Checking it for $\omega^{2}, k^{2} < m_{g}^{2}$ is insufficient. Our results imply that if one investigates this model exactly in the way it is written now (i.e. with the term $-\frac 1 {2 m_g^2} (\Box \phi)^2$), it does suffer from vacuum instability, and if we discard the higher derivative term at momenta greater that some cut-off, the instability becomes even worse. Is there any other way to save the new ekpyrotic scenario?

One could  argue \cite{Khoury,Creminelli:2005qk} that the term $\frac a {2 m_g^2} (\Box \phi)^2$ is just the first term in a sum of many higher derivative terms in an effective theory, which can be obtained by integration of high energy degrees of freedom of some extended physically consistent theory. In other words, one may conjecture that the theory can be made UV complete, and after that the problem with ghosts  disappears. However, not every theory with higher derivatives can be UV completed. In particular, the possibility to do it may depend on the sign of the higher derivative term  \cite{Adams:2006sv}.

According to \cite{Creminelli:2005qk}, the theory  (\ref{4d_L})  is
plagued by ghosts independently of the sign of the higher derivative term in the Lagrangian.  One can show it by introducing an
auxiliary scalar field $\chi$ and a new Lagrangian
\be \label{mixed}
{\cal L}' = -\sfrac12(\de \phi)^2 - a \: \de_\mu \chi \de^\mu \phi -
\sfrac12 a \: m_g^2 \: \chi^2 - V_{\rm int}(\phi)\; ,
\ee
which reduces exactly to ${\cal L}$ once $\chi$ is integrated out.  ${\cal
  L}'$ is diagonalized by the substitution $\phi = \phi' - a \chi$:
  \be \label{chi}
{\cal L}' = -\sfrac12 (\de \phi')^2 + \sfrac12
(\de \chi)^2 - \sfrac12 a \: m_g^2 \: \chi^2 - V_{\rm
  int}(\phi',\chi)\; ,
\ee
which clearly signals the presence of a
ghost: $\chi$ has a wrong-sign kinetic term.

 Then the authors of \cite{Creminelli:2005qk}
identified $\chi$ as a tachyon for $a=-1$, suggesting that in this case $\chi$ has
exponentially growing modes. However, this is not the case: due to the opposite sign of the kinetic term for the $\chi$-field, the tachyon is at $a=+1$, not at $a=-1$.  Indeed, because of the flip of the sign of the kinetic term for the field $\chi$, its equation of motion has a solution $\chi\sim  e^{\pm i (\omega t -\vec k\vec x) }$ with
\be
-(\omega^2-\vec k^2)= a\, m_g^2 \ .
\ee
For the field with the normal sign of the kinetic term, the negative mass squared  would mean exponentially growing modes.  But the flip of the sign of the kinetic term performed together with the flip of the sign of the mass term  does not lead to exponentially growing modes \cite{Linde:1984ir,Linde:1988ws}.
Based on the misidentification of the negative mass of the field with the wrong kinetic terms as a tachyon, the authors choose to continue with the $a=+1$ case in eq. (\ref{4d_L}). Starting from this point, their arguments are no longer related to the ghost condensate theory  and the new ekpyrotic theory, where $a = -1$. We will return to the case $a = -1$ shortly.

For the $a=1$ case  they argued that  the situation is not as bad as it could seem.
 They proposed  to use  the scalar
field theory eq.~(\ref{4d_L}) at energies below $m_g$, and
postulated that some new degree of freedom enters at $k > m_g$ and
takes care of the ghost instability.   The authors  describe this  effect by adding a term
$-(\de \chi)^2$ to construct the high energy Lagrangian.
For  $V_{\rm int} =0$ they postulate
\be \label{L_UV}
{\cal L}_{\rm UV}^{a=1} \equiv  {\cal L}'  -(\de \chi)^2 = -\sfrac12(\de \phi)^2 -\de_\mu \chi \de^\mu \phi
-(\de \chi)^2 - \sfrac12\: m_g^2 \: \chi^2 \;
\ee
and use  the shift $\phi=\tilde \phi-\chi$ to get a simple form of a UV theory.
This  trick reverses the sign of the kinetic term of the field $\chi$, and the ghost magically converts into a perfectly healthy scalar with mass $m_g$:
\be \label{L_UV1}
{\cal L}_{\rm UV}^{a=1} = -\sfrac12(\de \tilde \phi)^2
-\sfrac12(\de \chi)^2 - \sfrac12\: m_g^2 \: \chi^2 \; .
\ee

One may question validity of this procedure, but let us try to justify it by looking at the final result. Consider equations of motion for $\chi$ from eq.  (\ref{L_UV}) and solve them by iteration in the approximation when $\square \ll m_g^2$:
\be
\chi = {\square \phi\over m_g^2} +2 {\square \chi\over  m_g^2}\approx {\square \phi\over m_g^2}+ 2{\square^2 \phi\over m_g^4} +...
\label{sol} \ee
 Now replace $\chi$ in eq. (\ref{L_UV}) by its expression in terms of $\square \phi$ as given in eq.  (\ref{sol}). The result is our original Lagrangian   (\ref{4d_L}), plus some additional higher derivative terms, which are small at  $|\square| \ll m_g^2$, i.e. at $|\omega^{2}-k^{2}| \ll m_g^{2}$. Thus one may conclude that, for $a=1$, the theory (\ref{4d_L}), which 
has tachyonic ghosts, may be interpreted as a low energy approximation of the UV consistent theory (\ref{L_UV1}).

Now let us return to the ghost condensate/new ekpyrotic case. To avoid gradient instabilities in the ekpyrotic scenario, the sign of the higher derivative term in eq.  (\ref{4d_L}) has to be negative, $a=-1$, see eq. (1) and also eq. (\ref{Lp}) and the discussion below it. This means that  one should start  with eq. (\ref{4d_L}) with $a=-1$.

This theory is not tachyonic, but, as we demonstrated by performing its Hamiltonian quantization, it has ghosts, particles with negative energy, in its spectrum.  Can we improve the situation by the method used above? Let us start with the same formula,  ${\cal L}'  -(\de \chi)^2$, as in $a=+1$ case:
\be \label{L_UV-}
{\cal L}_{\rm UV}^{a=-1} \equiv  {\cal L}'  -(\de \chi)^2 = -\sfrac12(\de \phi)^2 +\de_\mu \chi \de^\mu \phi
-(\de \chi)^2 +\sfrac12\: m_g^2 \: \chi^2 \;
\ee
and replace $\chi$ by the iterative solution of its equation of motion
\be
\chi = {\square \phi\over m_g^2} - 2{\square \chi\over  m_g^2}\approx {\square \phi\over m_g^2}- 2{\square^2 \phi\over m_g^4} +...
\label{sol2} \ee
Thus, up to the terms which are small at $|\square| \ll m_g^2$, the theory with the Lagrangian (\ref{L_UV-}) does reproduce the model (\ref{4d_L}) with $a=-1$, up to higher order corrections in $|\square|/m_g^2$. The theory (\ref{L_UV-}) can be also  written as
\be \label{L_UV-1}
{\cal L}_{\rm UV}^{a=-1} = -\sfrac12(\de \tilde \phi)^2
-\sfrac12(\de \chi)^2 + \sfrac12\: m_g^2 \: \chi^2 \; .
\ee
where $ \phi= \tilde\phi+\chi$.
The sign of the kinetic term of both fields is normal, but the mass term still has the wrong sign, which leads to the tachyonic instability $\delta\chi \sim \exp \sqrt{m_g^2 - \vec k^2}\,t$.
Therefore  the cure for the ghost instability proposed in \cite{Khoury,Creminelli:2005qk}
does not  work for the case $a=-1$ of the ghost condensate/ekpyrotic scenario. 

Moreover, the procedure described above is valid only for $|\omega^{2}-\vec k^{2}| \ll m_g^{2}$. Meanwhile the gradient instability of the ekpyrotic theory in the regime of null energy condition violation ($c_{s}^{2} <0$) is most dangerous in the limit $\vec k^{2} \to \infty$, where this procedure does not apply, see (\ref{instgr}). 
This agrees with the general negative conclusion of Refs.   \cite{Adams:2006sv,Dubovsky:2006vk,Eling:2007qd,ArkaniHamed:2007ky} with respect to the theories of this type.

In this Appendix we analyzed the Lorentz-invariant theory (\ref{4d_L}) because the argument given in \cite{Khoury,Creminelli:2005qk} was formulated in this context. The generalization of our results for the ghost condensate/new ekpyrotic case is straightforward. Indeed, our results directly follow from the correlation between the sign of the higher derivative term in (\ref{mixed}) and the sign of the mass squared term in (\ref{chi}). One can easily verify that this correlation is valid independently of the value of $c_{s}^{2}$, i.e. at all stages of the ghost condensate/new ekpyrotic scenario.

To conclude, our statement that the ghost condensate theory and the new ekpyrotic scenario imply the existence of ghosts  is valid for the currently available versions of these theories, as they are presented in the literature. In this Appendix we explained why the  recent attempts  to make the theories with higher derivatives physically consistent  \cite{Khoury,Creminelli:2005qk} do not apply to the ghost condensate theory and the new ekpyrotic scenario. One can always hope that one can save the new ekpyrotic scenario and provide a UV completion of this theory in some other way, but similarly one can always hope that the problem of the cosmological singularity will be solved in some other way. Until it is done, one should not claim that the problem is already solved. 



\end{document}